# Forecasting the sustainable status of the labor market in agriculture.

**Malafeyev O.A.**
*Doctor of Physical and Mathematical Sciences, Professor, St. Petersburg State University, St. Petersburg, Russia*

**Onishenko V.E.**
*4th year student, St. Petersburg State University, St. Petersburg, Russia*

**Zaytseva I.V.**
*Candidate of Physical and Mathematical Sciences, Associate Professor, FGBOU VO «Stavropol State Agrarian University», Stavropol, Russia*

**Abstract**
In this article, a game-theoretic model is constructed that is related to the problem of optimal assignments. Examples are considered. A compromise point is found, the Nash equilibriums and the decision of the Nash arbitration scheme are constructed.

## INTRODUCTION

The economic demand on "educated" (high-skill) employees has been grown recently. The expected discounted salaries of high-skill employees have grown. Thus, the investments of time, effort and money into education have been attracting more and more citizens. The "educated employees" are supposed to be competent workers that have all the required skills and knowledge to accomplish their responsibilities. The education can reduce unemployment all over the world and Russia is not the exception. Education is an investment project with additional positive or negative benefit. The educational service is characterized by immateriality and variability; it's difficult for consumers to estimate its quality. It's also hard to compare this service with analogous services. These facts stipulated the complexity of labour force training research. The problem of labour force training research and necessity to enlarge the employees' level of education can be solved via economicmathematical methods. Assigning workers to the manufactures where the formers can get higher qualifications makes workers more capable to solve problems in their main area of employment. Mathematical models of such kind are studied in the papers [1-38].

For the case of six players, three models are built, for which there are given the solution of the assignments problem on the basis of different principles of optimality. Consider the set agricultural enterprises $H = \{h_1, \ldots, h_m\}$, which offer jobs and the set of agricultural workers $S = \{s_1, \ldots, s_n\}$, which wish to get a job at some agricultural enterprise. We assume that each agricultural enterprise $h_i$ has one vacancy. For this position the agricultural enterprise wishes to receive the employee. Each employee can be recruited for only one agricultural company. The worker can be assigned to a job at the enterprise only if an agreement is reached between him and the agricultural enterprise. The result of the game is the appointment of an employee to agricultural enterprise.

### Model 1.

Consider the game in normal form. [5]
$$\Gamma = <I, \{X\}_{i=1}^{6}, \{H\}_{i=1}^{6}>.$$

Here
- $I = \{1,2,3,4,5,6\}$ is a set of players. Players with numbers 1,2,3 belong to the sub-set $S$ of the set $I$. Players with numbers 4,5,6 belong to the subset $H$.
- $X_i$ is the set of strategies of the player $i$. $X = \prod_{i=1}^{6} X_i$ is the set of 6-tuples (possible situations) in the game.
- $H_i: X = \prod_{i=1}^{6} X_i \to R_1$ —is a payoff function of the player $i$.

Each player has three strategies. The player from the set $S$ can choose only one player from the set $H$. . The player from the set $H$ can choose only one player from the set $S$.

Let's rename the players from the subset as follows: $H = \{h_1, h_2, h_3\}$, implying that $h_1$ is the player under number 1 (in the set $H$), $h_2$ the player under the number 2 and $h_3$ is the player under the number 3. Formally, every assignment of a worker to a work can be represented as substitution of the form:

$$\begin{pmatrix} 1 & 2 & 3 \\ h_k & h_l & h_m \end{pmatrix}.$$

The first line is unchanged here. It corresponds to the workers' numbers, the second line corresponds to the jobs. In our problem there are six such substitutions:

$P = \{p_1, p_2, \ldots, p_6\}.$

$$p_1 = \begin{pmatrix} 1 & 2 & 3 \\ h_1 & h_2 & h_3 \end{pmatrix},$$

$$p_2 = \begin{pmatrix} 1 & 2 & 3 \\ h_2 & h_1 & h_3 \end{pmatrix},$$

$$p_3 = \begin{pmatrix} 1 & 2 & 3 \\ h_3 & h_2 & h_1 \end{pmatrix},$$

...

$$p_6 = \begin{pmatrix} 1 & 2 & 3 \\ h_2 & h_3 & h_1 \end{pmatrix}.$$

Similarly, we get six substitutions from the set $Q = \{q_1, \ldots, q_6\}$. They correspond to the distribution of jobs for workers. The first line corresponds to the number of jobs in the set $H$. The second line of the substitution corresponds to workers under the numbers 1,2 and 3. These are the substitutions of the form:

$$\begin{pmatrix} 1 & 2 & 3 \\ s_k & s_l & s_m \end{pmatrix}.$$

A situation in the game $\Gamma$ viewed as substitution. Note that in this way we do not exhaust the set of all situations in the game. We distinguish only those of them, which make it possible to make substitutions.

It is necessary to find the full compromise set $C_H$ and the set $N$ of Nash equilibrium situations.

We introduce utility matrices $A, B$ for players from subsets $S$ and $H$:
Matrix A (the rows correspond to the numbers of players from subsets of $S$, and the columns correspond to the set $h_1, h_2, h_3$ respectively.)

$$A = \begin{pmatrix} 75 & 22 & 94 \\ 33 & 41 & 86 \\ 45 & 13 & 54 \end{pmatrix}$$

Matrix B (the rows correspond to the set $h_1, h_2, h_3$ respectively, and the columns correspond to ordinal numbers of players from . Then

$$B = \begin{pmatrix} 94 & 71 & 17 \\ 30 & 32 & 18 \\ 59 & 85 & 38 \end{pmatrix}.$$

We define the players winning functions on the substitution set $P$ as follows:

$$H_1(p_1) = \alpha_{1h_1} = 76,$$

$$H_2(p_1) = \alpha_{2h_2} = 41,$$

$$H_3(p_1) = \alpha_{3h_3} = 54,$$

$$H_4(p_1) = \beta_{h_11} = 94,$$

$$H_5(p_1) = \beta_{h_22} = 32,$$

$$H_6(p_1) = \beta_{h_33} = 38,$$

$\alpha_{lh_k}$ – is the element of matrix $A$ at the intersection $l$-ой row and $h_k$-th column, and $\beta_{h_kl}$ – is an element of the matrix $B$ at the intersection $h_k$-th row and the $l$-th column. On the set of substitutions $Q$ the winning functions are defined in a similar way. On substitutions from sets $P$ and $Q$ within identical numbers, the payoff functions will coincide.

Substitutions with identical numbers correspond to the same situation in the game. Therefore, the number of different situations in the game is two times less than the total number of substitutions and equal to six (we take this into account when drawing up a payoff matrix). The matrix has the form( the rows correspond to substitutions $p_1, \ldots, p_6$, which form the set of situations X, the columns correspond to the numbers of players from the set $I$):

$$\begin{pmatrix} 76 & 41 & 54 & 94 & 32 & 38 \\ 22 & 33 & 54 & 30 & 71 & 38 \\ 94 & 41 & 45 & 59 & 32 & 17 \\ 94 & 33 & 13 & 94 & 71 & 18 \\ 76 & 86 & 13 & 94 & 85 & 18 \\ 22 & 86 & 45 & 30 & 85 & 17 \end{pmatrix}$$

We calculate the real vector $M = (M_1, \ldots, M_6)$ with the formula:

$$M_i = \max_{x \in X} H_i(x)$$

We obtain the following result:

$$M = (94, 86, 54, 94, 85, 38).$$

We compute the compromise set $C_H$ with the formula:

$$C_H = \{x \in X : \max_i (M_i - H_i(x)) \leq \max_i (M_i - H_i(x')), \forall x' \in X\}$$

We obtain the following result:

$$C_H = \{p_5\}.$$

The situation $p_5$ leads to the following assignments: the first worker gets assigned to the first enterprise (the fist enterprise get the first worker), the second one gets assigned at the third enterprise, and the third one at the second. Notice, that the full compromise set is such that the least satisfied player gets a guaranteed prize. In accordance with the winning function the third player is satisfied less than the rest in this situation and gets their guaranteed prize equal to 13, or $H_3(p_5) = 13$.

The set of Nash equilibrium situations is found thus:

$$\{x^* \in X : H_i(x^* \parallel x) \leq H_i(x), \forall i \in I\},$$

$$\phi \varphi \chi(x^* \parallel x) = (x_1, \ldots, x_{i-1}, x_i^*, x_{i+1}, \ldots, x_6).$$

In our problem, if any of the six players changes their decision, while the rest do not, we cannot complete the assignment. In other words ,if the winning function equals to zero. Therefore all the situations in the game are examples of Nash equilibrium.

**Model 2.**

Let $H = (h_1, h_2, h_3)$—be the set of job and,
$S = (s_1, s_2, s_3)$— be the set of workers. The efficiency matrices are set for both workers and jobs, respectively. We will solve two optimal assignment problems: assign workers to jobs and distribute the jobs among workers so as to maximize the overall utility for both.
 1. Assignment problem for workers.

Matrix
$$A = \begin{pmatrix} 40 & 20 & 10 \\ 15 & 12 & 8 \\ 32 & 30 & 18 \end{pmatrix}$$

where $a_{ij}(i = 1..3, j = 1..3)$ — is the efficiency of assigning a worker to a job. We solve the problem using the Hungarian algorithm and get the assignment matrix.

$$X = \begin{pmatrix} 1 & 0 & 0 \\ 0 & 0 & 1 \\ 0 & 1 & 0 \end{pmatrix}$$

Then the effectiveness of assignments will be

$$\sum_{i=1}^{3} \sum_{j=1}^{3} a_{ij} * x_{ij} = 40 + 8 + 30 = 78$$

2. The problem of the distribution of jobs among workers.
    Matrix
$$B = \begin{pmatrix} 9 & 14 & 21 \\ 11 & 7 & 5 \\ 8 & 16 & 25 \end{pmatrix}.$$

where $b_{ij}(i = 1..3, j = 1..3)$ — is the efficiency of the distribution of jobs for the worker. Solving the problem in a similar way, we obtain the assignment matrix.

$$Y = \begin{pmatrix} 0 & 0 & 1 \\ 1 & 0 & 0 \\ 0 & 1 & 0 \end{pmatrix}.$$

Then the distribution efficiency

$$\sum_{i=1}^{3} \sum_{j=1}^{3} b_{ij} * y_{ij} = 21 + 11 + 16 = 48.$$

From the first and second tasks it is clear that two of the three appointments did not coincide. Thus, if we assign workers to jobs and distribute the jobs among workers independently of each other, maximizing only our own effectiveness, then it is not possible to make acceptable assignments. One solution to this problem is to create a union of workers and a trade union of employers who, caring about the interests of their members, seek to find a mutually beneficial solution through negotiations.

### Model 3.

The task that this model solves is to come to an agreement for reasonable players in the joint choice of the solution in the course of negotiations. Set a bimatrix game for two players - the S workers 'union and the employers' union H the payoff matrix:

$$\begin{pmatrix} (6,2) & (0,0) \\ (0,0) & (2,6) \end{pmatrix}$$

On the plane $(K_1, K_2)$ ($K_1$ is the payoff function of the workers 'union (player 1), $K_2$ is the payoff function of the employers' union (player 2)), consider the set R corresponding to the possible payoff vectors in joint mixed strategies for our game. Together, players can realize any gain in mixed strategies in the R area (the boundary of the R-triangle with the vertices $O(0,0), a(2,6), b(6,2)$). However, this does not mean that they can agree on any outcome of the game. So, for union S the point (6,2) is most preferable, and for trade union H-point (2,6) ) is most preferable. None of the players will agree with the results of the negotiations if his payoffs are less than the maximum value, since he can get this payoff independently (regardless of the partners actions ). Maximine mixed strategies are as follows:

$$\max_x \min\{K_1(x, \beta_1), K_1(x, \beta_2)\} = \min\{K_1(x^0, \beta_1), K_1(x^0, \beta_2)\}$$

where $\beta_1, \beta_2$ are the player's pure strategies 2, $x^0 = (\xi^0, 1 - \xi^0), 0 \leq \xi \leq 1$ is the mixed strategy of player 1, which he intends to select in order to maximize one of the two values $K_1(x, \beta_1), K_1(x, \beta_2)$.
Maximin mixed strategies of players have the form
$x^0 = (\frac{1}{4}, \frac{3}{4})$, $y^0 = (\frac{3}{4}, \frac{1}{4})$ , respectively, and the payoff vector in the maximin strategies $(v_1^0, v_2^0) = (\frac{3}{2}, \frac{3}{2})$. Therefore, the set S, negotiable, is bounded by the points $a(2,6), b(6,2), c, d(\frac{3}{2}, \frac{3}{2}), e$, e (c, e are the intersection points of the rays emerging from the point d and parallel to the axes, with the sides of the triangle). This is the negotiating set of the game. Further, acting together, players can always agree to select points on the segment $\overline{ab}$, since it is beneficial to both of them (the segment $\overline{ab}$ corresponds to the Pareto optimal situations). It is necessary to choose a point $(\overline{v_1}, \overline{v_2})$ from the set S, which will be obtained in the result of negotiations. The Nash arbitrage scheme [7] gives the vector $(\overline{v_1}, \overline{v_2}) = (4,4)$

**Acknowledgement**
The work is partially supported by the RFBR grant No. 18-01-00796